\documentclass[prc,aps,a4paper,groupedaddress,superscriptaddress,nofootinbib,showpacs
,preprintnumbers,twocolumn]{revtex4}
\usepackage{graphicx}
\usepackage{amsfonts}
\usepackage{amssymb}
\usepackage{natbib}
\usepackage{dcolumn}
\usepackage{bm}
\newcommand{\bwt}{\begin{widetext}}
\newcommand{\ewt}{\end{widetext}}
\newcommand{\beq}{\begin{equation}}
\newcommand{\eeq}{\end{equation}}
\newcommand{\bea}{\begin{eqnarray}}
\newcommand{\eea}{\end{eqnarray}}
\begin{document}
\title{Self-consistent random phase approximation in the Sch\"utte-Da-Providencia 
fermion-boson model}
\author{A. Rabhi}
\email{rabhi@ipnl.in2p3.fr}
\affiliation{Laboratoire de Physique de la Mati\`ere Condens\'ee, 
Facult\'e des Sciences de Tunis, Campus Universitaire, Le Belv\'ed\`ere-1060, Tunisia} 
\affiliation{IPN-Lyon, 43 Bd du 11 novembre 1918, F-69622 Villeurbanne Cedex, France}
\date{\today}
\begin{abstract}
The Self-consistent random phase approximation~(SCRPA) is applied to the exactly 
solvable model with fermion boson coupling proposed by Sch\"utte and Da-Providencia. 
Very encouraging results in comparison with the exact solution of the model for 
various observables are obtained. The transition from the normal phase to the phase 
with a spontaneously broken symmetry is carefully investigated. The strong reduction 
of the variance in SCRPA vs HF is pointed out. 
\end{abstract}
\maketitle

During the last decade the so-called Self-Consistent version of the Random Phase Approximation
(SCRPA) has seen very encouraging successes in a number of non trivial model cases (see for
example~\cite{b1} for a detailed description of the method and~\cite{b2} for the application of SCRPA
to the many level pairing model). In spite of these performances of the theory, there are
remaining problems. In first place this concerns situations with spontaneously broken
symmetries. Such situations were treated in~\cite{b3, b4, b1}. Whereas in the Lipkin model~\cite{b3} the
symmetry broken ('deformed') phase caused no problem because the broken symmetry is discrete
(parity), in the other two cases~\cite{b4,b1}, with a continuous broken symmetry, problems appeared
with the low-lying mode known to be exactly at zero energy in the standard HF-RPA approach (the
spurious or Goldstone mode). In the two cases cited~\cite{b4, b1} the low-lying mode does not appear
at zero energy in SCRPA because the RPA operator does not contain the symmetry operator as a
limit case. Indeed, \textit{e.g.} the number operator in quasiparticle (BCS) representation
contains a purely hermitian piece $\alpha^{\dagger}_{k}\alpha_{k}$ which can not be incorporated 
in the RPA operator which by definition is non hermitian. The same situation is present in the
Sch\"utte-Da-Providenica boson-fermion model~\cite{b5} where the symmetry operator contains the boson 
and fermion number operators. The violation of the Goldstone theorem signifies that the Ward
identities and conservations laws are not respected. Though this violation seems relatively mild
and to go away in macroscopic systems (the hermitian pieces becoming of zero weight), the
situation remains annoying for finite systems.

In this note which can be considered as a sequel of~\cite{b4} we want again to investigate 
the Sch\"utte-Da-Providenica model 
\beq
H=\bar{n}+\alpha b^{\dagger} b + G (\tau^{+} b^{\dagger} +\tau^{-} b)
\label{a1}
\eeq
with $b^{\dagger}$, $b$ ideal boson operators and,
$n=\sum_{i=1}^{N}a^{\dagger}_{0i}a_{0i}$, $\bar{n}=\sum_{i=1}^{N}a^{\dagger}_{1i}a_{1i}$,
$\tau^{+}=\sum_{i=1}^{N}a^{\dagger}_{1i}a_{0i}$, $\tau^{-}=\sum_{i=1}^{N}a^{\dagger}_{0i}a_{1i}$, 
$\tau^{0}=\frac{1}{2}{(\bar{n}-n)}$,
where the $a^{\dagger}$, $a$ are fermion operators.
In analogy to the work in~\cite{b4} we will introduce the more general RPA operator
\beq
Q^{\dagger}_{\nu}=X_{\nu}t^{+}-Y_{\nu}t^{-}+\lambda_{\nu}B^{\dagger}-\mu_{\nu}B
+U_{\nu}\beta^{\dagger}\beta^{\dagger}-V_{\nu}\beta\beta, \nu =1,2,3
\label{a3}
\eeq
where
\beq
t^{\pm}=\frac{T^{\pm}}{\sqrt{-2\langle T^{0}\rangle}}\quad \text{and}\quad 
\beta^{\dagger}\beta^{\dagger}=\frac{B^{\dagger}B^{\dagger}}
{\sqrt{2(1+2\langle B^{\dagger}B\rangle)}}.
\eeq
The operators $T^{\pm}$,~$T^{0}$ are obtained from $\tau^{\pm}$,~$\tau^{0}$ 
by writing the latter ones in the "deformed" basis
\begin{equation}
\pmatrix{\alpha^{\dagger}_{1k} \cr \alpha_{0k}}=
\pmatrix
{u &-v \cr v & u} \pmatrix{a^{\dagger}_{1k} \cr a^{\dagger}_{0k}},\quad u^{2}+v^{2}=1.
\label{a5}
\end{equation}
The bosons operators $B^{\dagger}$ and $B$ are obtained from the original ones 
by a shift transformation $B\rightarrow b-\sigma$,
where $\sigma$ is a $c$-number characterizing the appearance of the Bose condensate.
The introduction of the boson pair operators~$\beta^{\dagger}\beta^{\dagger}$ is motivated by
the fact that otherwise there exists a certain dissymmetry between fermions and bosons, the
fermions being in any case bilinear whereas the bosons are otherwise only contained to linear order
in~(\ref{a3}). Also the symmetry operator $P=b^{\dagger}b-\bar{n}$ contains the bosons
quadratically and the extended ansatz~(\ref{a3}) may therefore show improved behavior with
respect to the Goldstone mode. The formalism goes exactly in the same way as in~\cite{b1,b2,b4} 
using the equation of motion method 
\beq
\langle [\delta Q, [H^{\prime}, Q^{\dagger}_{\nu}]]\rangle = 
\Omega_{\nu}\langle [\delta Q, Q^{\dagger}_{\nu}]\rangle
\label{a7}
\eeq  
to determine the amplitudes in~(\ref{a3}). As in~\cite{b4}, in order to fix the value
$L=\langle P\rangle$, we use in~(\ref{a7}) the cranked Hamiltonian $H^{\prime}=H-\mu P$ 
in the symmetry broken phase, otherwise $H^{\prime}=H$. The mean-field amplitudes $u$, 
$v$ and $\sigma$ are readily obtained from a minimization of the ground-state energy, 
leading to $\langle [H^{\prime}, t^{+}]\rangle=\langle [H^{\prime},B^{\dagger}]\rangle=0$. 
The amplitudes in~(\ref{a3}) form a complete orthonormal set when calculated 
from~(\ref{a7}). Then~(\ref{a3}) can be inverted and with the usual condition for 
the RPA ground-state 
\beq
Q_{\nu}|\text{RPA}\rangle =0, \quad\nu =~1,~2,~3.
\label{a8}
\eeq
all expectation values appearing in~(\ref{a7}) like for example $\langle t^{+}B^{\dagger}\rangle$,
$\langle t^{+}B\rangle$ and $\langle B^{\dagger}B\rangle$ can directly be expressed in terms of the
RPA amplitudes. The only unknown quantity at this point preventing a fully self-consistent 
solution of the SCRPA equations~Eq.~(\ref{a7}) is the expectation value 
$\langle T^{0}\rangle$. However, in analogy to our previous study for the two-level pairing 
model~\cite{b1} this quantity can be expressed as an expression in the operators $T^{+}$ and 
$T^{-}$ up to any order in a fast converging series according to
\beq
T^{0}=-\frac{N}{2}+\frac{1}{N} T^{+}T^{-} 
+ \frac{1}{N^{2}(N-1)}{T^{+}}^{2}{T^{-}}^{2} + \ldots.
\eeq
With this relation the SCRPA equations are completely closed and we can proceed to the numerical 
solution. We notice that with respect to~\cite{b4} the SCRPA Eq.~(\ref{a7}) is a $6\times 6$
dimensional problem whereas before it was $4\times 4$.

We now come to the presentation and discussion of the results. In what follows, 
SCRPA(6) refers to the SCRPA method with RPA excitation operator quadratic in 
the bosons operators \textit{i.e.}~Eq.~(\ref{a3}), while SCRPA(4) refers to the same method but 
with RPA excitation operator linear in the bosons operators \textit{i.e.}~$U_{\nu}=V_{\nu}=0$. 
In the following we also use the set of parameters $\alpha =3$, $N = 30$ as in~\cite{b4}, 
for which in this model the phase transition point is localized at $x=1.0$, where $x=G\sqrt{N/\alpha}$. 
SCRPA always shows a clear superiority over standard RPA, though, besides for some quantities, the
differences are not very pronounced. Concerning the ground-state energy we do not give 
results but we only notice that we arrive practically at the same interpretations as in~\cite{b4}. 
However, in order to test the accuracy of our approach it is instructive to calculate 
the differences of energies of the ground-state band with $L$ values just one unit away from the 
absolute ground-state. One such quantity is the "chemical potential" which should be identified 
with the Lagrange parameter used for restoring the symmetry 
\beq
\mu=\frac{1}{2}\left(E^{0}_{L+1}-E^{0}_{L-1}\right).
\label{mu}
\eeq   
In Table~\ref{tab1} we show $\mu$ when we calculate separately $E^{0}_{L \pm 1}$ 
(in standard RPA and SCRPA) and then take the difference. We also give 
in Table~\ref{tab1} the $L$ values which correspond for a given $x$ value to the 
absolute ground-state. In Table~\ref{tab1} we see a strong improvement of SCRPA(6) 
and SCRPA(4) over standard RPA and the high quality of the results in comparison with 
the exact values in the region around the phase transition point. Also SCRPA(6) is still 
improved over SCRPA(4). We could also have taken the $\mu$ values found from adjusting the 
correct $L=\langle P\rangle$ values in the standard RPA and SCRPA calculations; we have 
checked numerically that the results are always practically identical. 
\begin{table}[htb]
\begin{tabular}{cccccc}
\hline
\hline
~$x$~&~$L$~&~$\mu(\text{exact})$~&~$\mu(6)$~&~$\mu(4)$~&~$\mu(\text{RPA})$~\\ \hline
 1.1 & -1  &    -0.0131  & -0.0485          & -0.0485  &  0.1160   \\
 1.4 & -3  &     0.0325  &  0.0348          &  0.0405  &  0.1303   \\
 1.8 & -3  &    -0.0341  & -0.0315          & -0.0329  &  0.0060   \\
 2.2 &  0  &     0.0320  &  0.0300          &  0.0240  &  0.0345   \\
 2.6 &  4  &     0.0305  &  0.0285          &  0.0221  &  0.0207   \\
 3.0 &  9  &     0.0204  &  0.0195          &  0.0135  &  0.0071   \\ 
\hline
\end{tabular}
\caption{Chemical potential : exact, SCRPA(6), SCRPA(4) and standard RPA; $L$ 
values, in the deformed region for different values of the interaction strength $x$.}
\label{tab1}
\end{table}     

\begin{figure}[htb]                       
\centering
\vspace{0.3 in}
\includegraphics[width=0.9\linewidth]{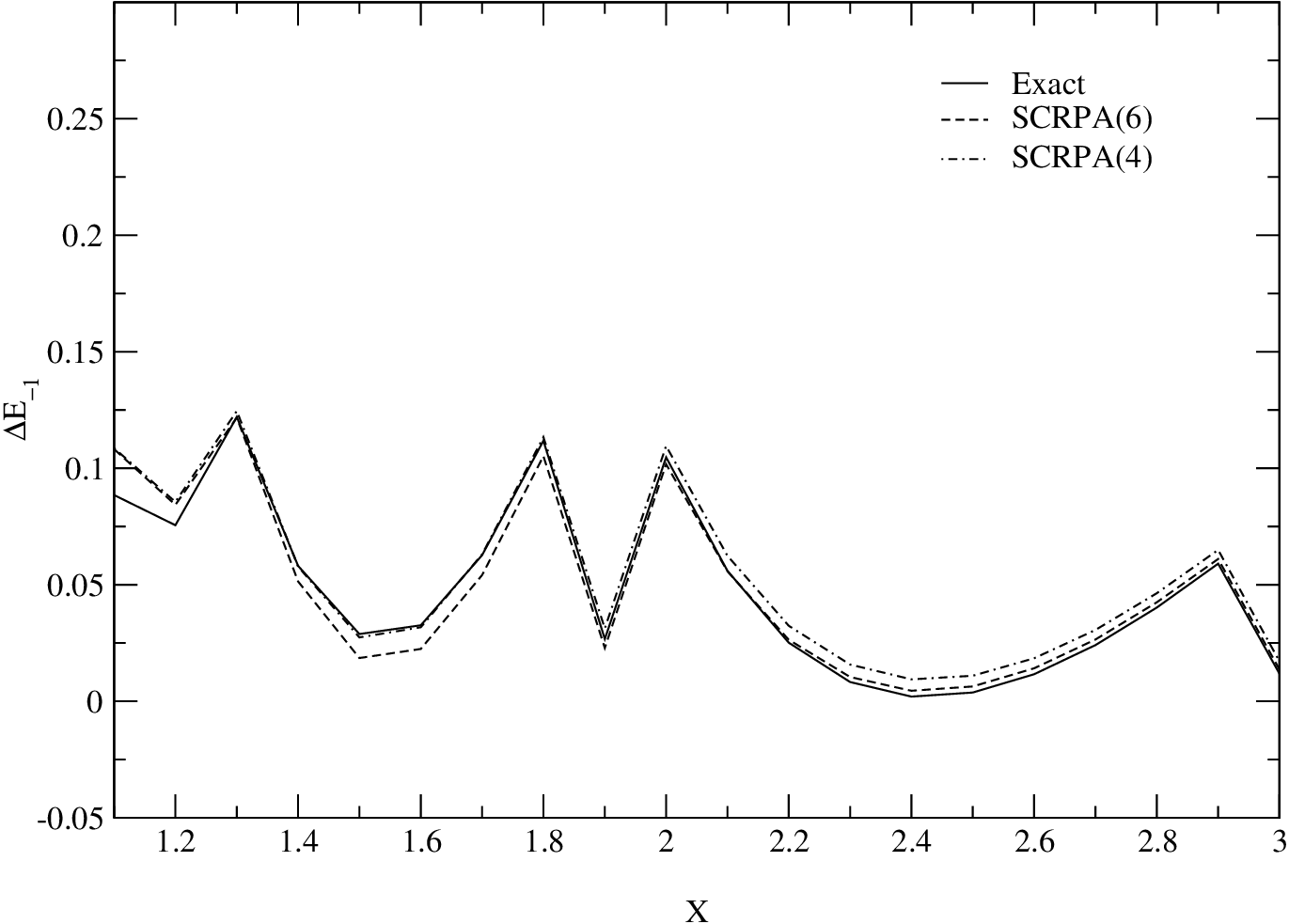} 
\caption{Comparison between the exact, SCRPA(6) and SCRPA(4) results for the excitation
energy $\Delta E_{-1}=E^{0}_{L}-E^{0}_{L-1}$ in the deformed region.}
\label{minus}
\end{figure}

Two other interesting quantities to be calculated within the SCRPA formalism and 
closely related to the chemical potential are the energy differences of the absolute 
ground-state with its "left" and "right" neighbors just one unit in $L$ away 
\beq
\Delta E_{\pm 1}=\pm\left(E^{0}_{L \pm 1}-E^{0}_{L}\right). 
\eeq
These quantities are interesting because, as we will explain below, they should be closely 
related to the lowest RPA eigenvalue $\Omega_{1}$ in the symmetry broken phase. 
Because we obtain similar interpretation and conclusions for both $\Delta E_{\pm 1}$, 
we will present and discuss only the result for $\Delta E_{- 1}$. 
In Fig.~\ref{minus} we see a very good agreement with the exact results of both SCRPA(6) 
and SCRPA(4). However, we note that in this quantity no clear superiority of SCRPA(6) 
over SCRPA(4) can be detected, the results being at times in favor of the one or the other.
The good quality of the results for $\Delta E_{\pm 1}$ shows 
that the SCRPA method is able to reproduce the full spectrum. We also should notice that the 
smallness of $\Delta E_{\pm 1}$ means that two neighboring ground-states with $L$ and $L \pm 1$, 
respectively, are almost degenerate which indicates that the system is in the phase of 
spontaneously broken symmetry. Furthermore, one can check that in the large $N$ limit 
$\Delta E_{\pm 1}$ tends to zero. The zero eigenvalue (Goldstone mode) which is one of the 
solutions of the standard RPA in the deformed region corresponds to this degeneracy of the neighboring 
ground-state energies in the large $N$ limit.          

Let us now discuss the eigenvalues of RPA and SCRPA matrices. As it is well known~\cite{b5, b4}, 
in standard RPA the lowest eigenvalue corresponds in the symmetry broken (deformed) region to 
the spurious mode $\Omega_{1} =0$, whereas the second eigenvalue gives the excitation of the 
intrinsic system. Before coming to this point we should mention again that the RPA eigenvalues 
in the deformed region are calculated with the "intrinsic" Hamiltonian $H^{\prime}=H-\mu P$. 
Therefore, when the symmetry is restored due to the appearance of the Goldstone mode, the RPA 
eigenvalues give the excitation energies of the system. The results for the mode $\Omega_{2}$ 
are not shown in a figure because we obtain the same interpretation as already 
given in~\cite{b4}, however, with still improved results from SCRPA(6).      

\begin{figure}[htb]                       
\centering
\vspace{0.3 in}
\includegraphics[width=0.9\linewidth]{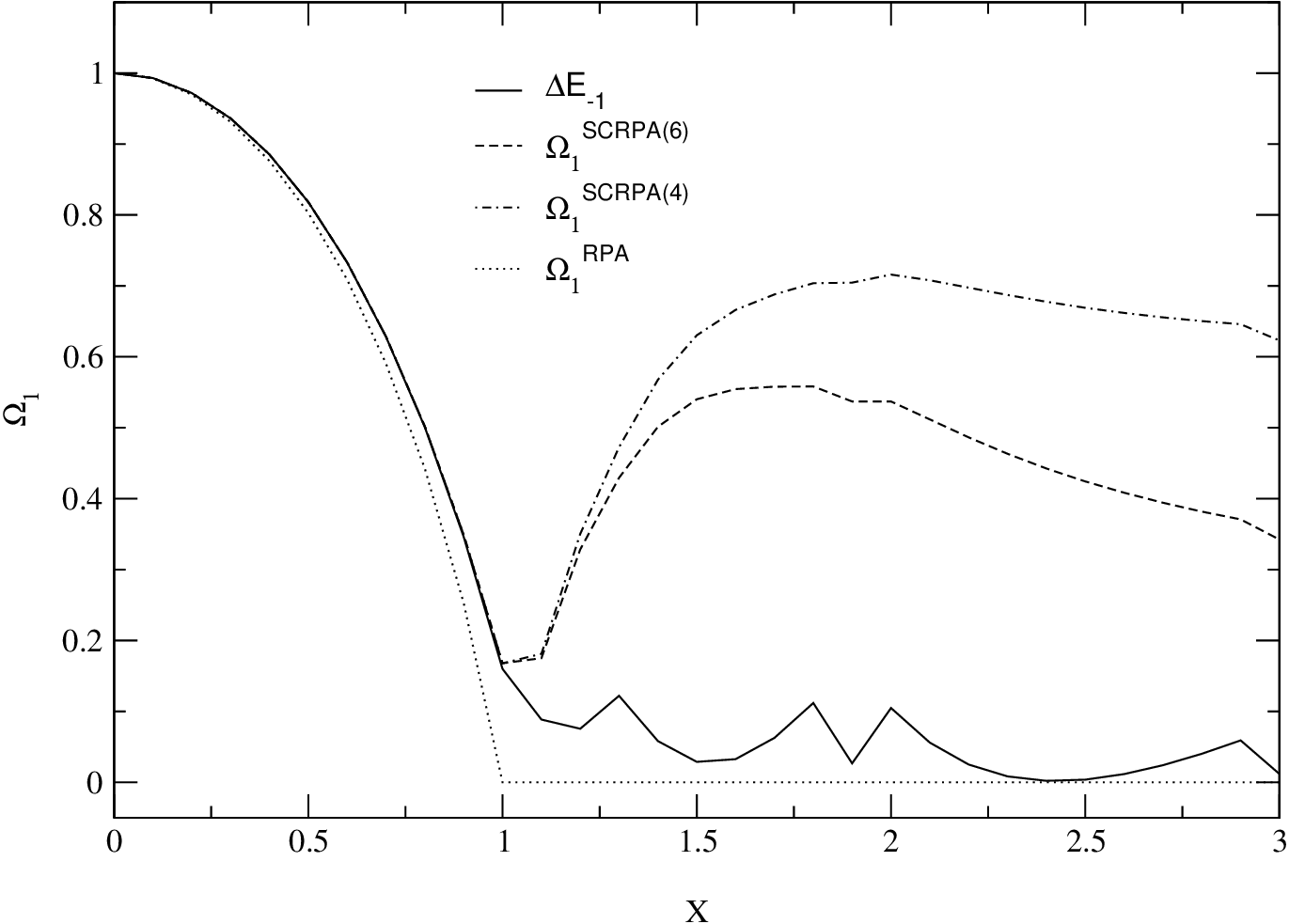}
\caption{The standard RPA, SCRPA(4) and SCRPA(6) results for the spurious mode 
$\Omega_{1}$ compared with the exact energy of the excitation $\Delta E_{-1}$.}
\label{omega1}
\end{figure}

Concerning the low-lying eigenvalue of SCRPA which in standard RPA 
corresponds to the zero-energy eigenvalue (Goldstone or spurious mode) in 
the "deformed" region, we present in Fig.~\ref{omega1} a comparison 
between standard RPA, SCRPA(4) and SCRPA(6) with exact results. 
In the "spherical" phase we notice that the eigenvalue $\Omega_{1}$ 
is identified with the exact "intraband" excitation $\Delta E_{-1}$. 
Furthermore, we see the important improvement of the SCRPA results in 
both cases SCRPA(4) and SCRPA(6) with respect to standard RPA result. 
After the phase transition $\Delta E_{-1}$ remains finite but very small, 
slowly decreasing for increasing $x$, while the lowest eigenvalue in standard RPA 
corresponds to the spurious mode $\Omega_{1}=0$. Concerning the low-lying eigenvalue in SCRPA 
calculation we see that SCRPA(6) improves the result with respect to SCRPA(4) 
but it is still quite far from the exact result. We can notice that the Goldstone theorem 
is not correctly fulfilled in this case. Therefore, the problem of the identification of 
the low-lying eigenvalue is not yet solved in SCRPA in spite of the introduction of the 
quadratic boson terms in~(\ref{a3}).

\begin{figure}[htb]                       
\centering
\vspace{0.3 in}
\includegraphics[width=0.9\linewidth]{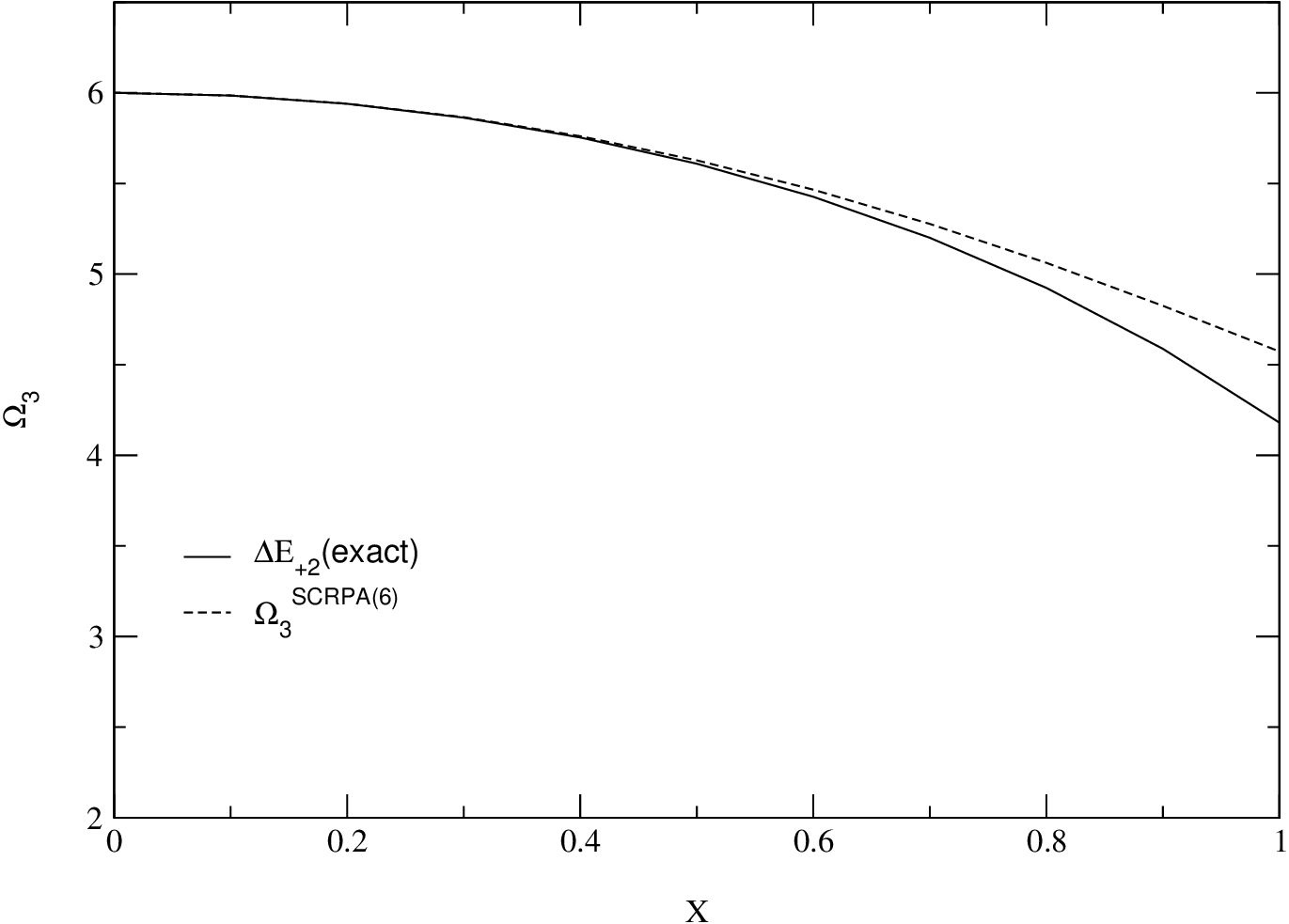}                      
\caption{The SCRPA(6) results of the energy of the excitation mode $\Omega_{3}$ 
compared with the exact energy of the excitation $\Delta E_{+2}$ in the spherical 
region.}
\label{omega3}
\end{figure}

From the presence of the quadratic boson terms in the RPA excitation operator the SCRPA 
method produces an supplementary eigenvalue which is noted $\Omega_{3}$. 
Let us now discuss the results for this mode, \textit{i.e.} the third eigenvalue 
RPA of the SCRPA(6) which is presented in Fig.~\ref{omega3}. In the spherical region 
we notice that this mode is identified with very good accuracy to the exact "intraband" 
excitation $\Delta E_{+2}$. However, in the deformed region (not shown), the result obtained for 
this mode cannot be identified with one of the exact excitations of the system.
At present we do not have an explanation of this fact. It is likely related to the failure of the
Goldstone theorem mentioned above. See also further discussion of this point at the end of this
note. The difficulty may be of the same origin as with the $\Omega_{1}$ mode. 

\begin{figure}[htb]                       
\centering
\vspace{0.3 in}
\includegraphics[width=0.9\linewidth]{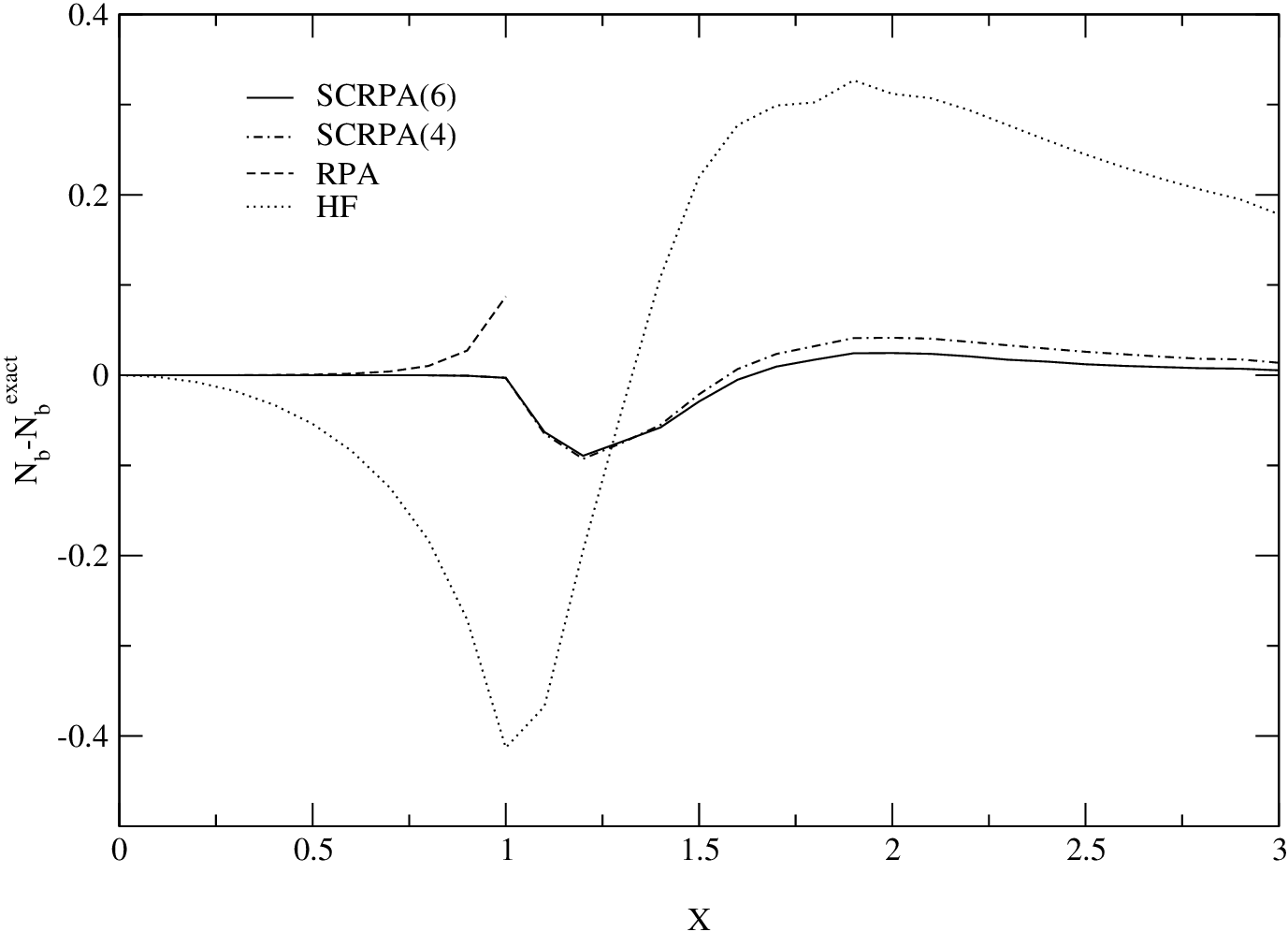} 
\caption{The difference $[N_{b}-N^{\text{exact}}_{b}]$ calculated with SCRPA(6), SCRPA(4),
RPA and mean field methods as function of the interaction strength $x$.}
\label{number}
\end{figure}

A quantity which is particularly sensitive to the correct treatment of correlations 
in the ground-state is the mean boson number~(not shown in Ref.~\cite{b4}). 
This expectation value can be obtained in terms of the RPA amplitudes according to
\beq
N_{b}=\langle b^{\dagger}b \rangle = \langle B^{\dagger}B\rangle +\sigma^{2},
\eeq
where $\langle B^{\dagger}B\rangle$ is given, in SCRPA(6), by 
$\langle B^{\dagger}B\rangle =\mu^{2}_{1}+\mu^{2}_{2}+\mu^{2}_{3}$.
In Fig.~\ref{number} we show the results of the SCRPA(6), SCRPA(4), standard RPA and 
mean field methods for this quantity. Again with SCRPA(6) one notices a significant improvement 
over standard RPA and HF method for which the agreement with the exact result is not satisfying. 
Also, we note in Fig.~\ref{number} that the SCRPA(6) improves slightly the result over the SCRPA(4) 
specially in the deformed region. Furthermore, we note that in standard RPA method, because 
we have a Goldstone mode in the deformed region, we cannot calculate this quantity. 

\begin{figure}[htb]                       
\centering
\vspace{0.4 in}
\includegraphics[width=0.9\linewidth]{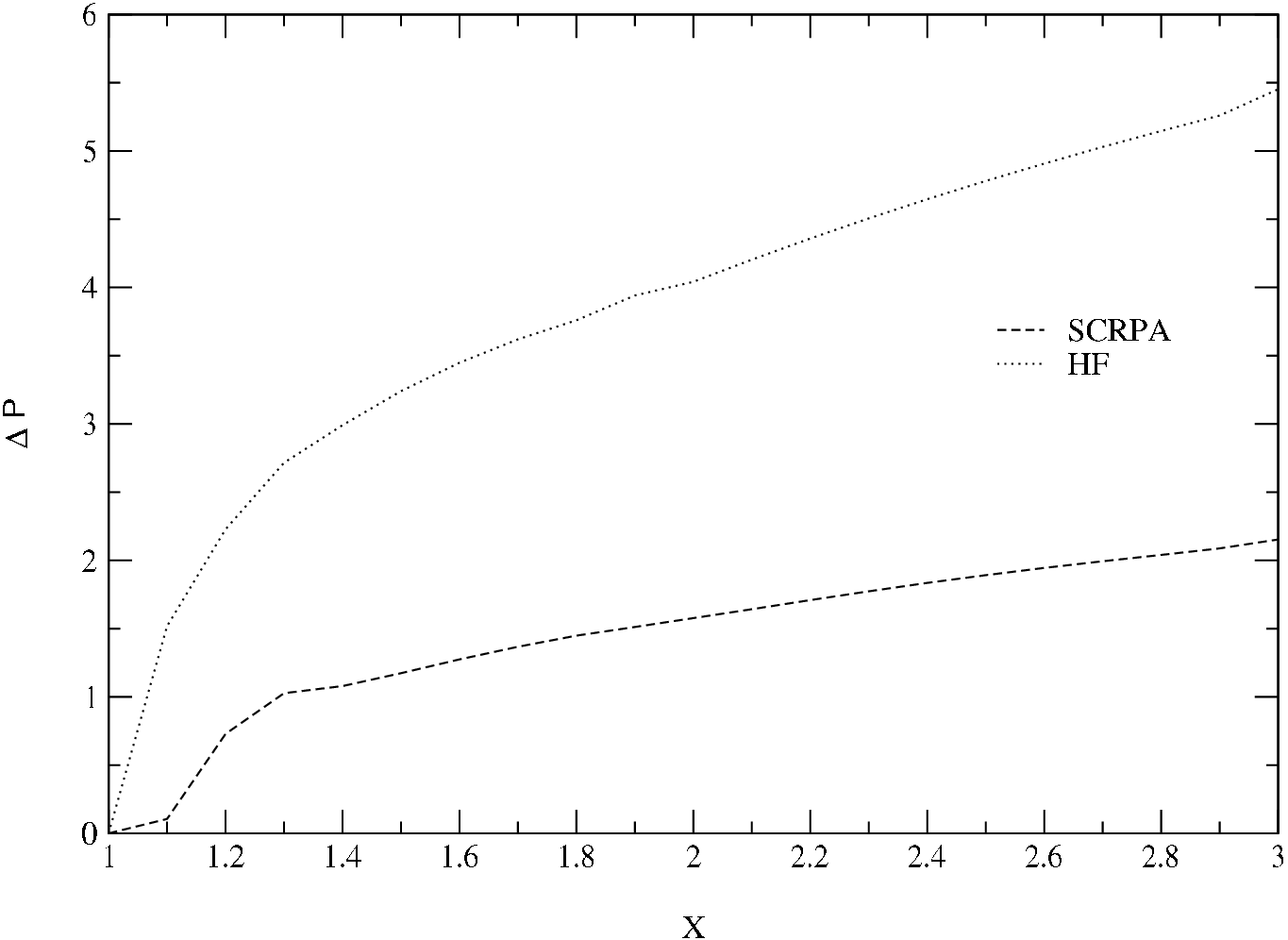} 
\caption{Variance as function of the interaction strength $x$.}
\label{fluct}
\end{figure}

Another quantity which is very interesting to investigate in the SCRPA method 
is the variance of the symmetry operator $P$ given by 
\beq
{\Delta P}^{2} = \langle P^{2}\rangle - {\langle P\rangle}^{2}.
\eeq
In Fig.~\ref{fluct} we present the results corresponding to this quantity calculated 
with SCRPA and HF methods. This is a new result which was not elaborated in~\cite{b4}. 
We see that the variance is strongly reduced compared to HF values. 
We, however, see that $\Delta P$ even in the SCRPA acquires sizable non vanishing values. 
This simply means that the symmetry $P$, broken at the level of the mean field theory, 
is not completely restored. Furthermore, we do not present the standard RPA results concerning 
this quantity in Fig.~\ref{fluct} because the RPA amplitudes
originating from the Goldstone mode are divergent. This constitutes the same kind of situation 
as that for the boson number in the ground-state calculation for which we also have not 
given the standard RPA results in deformed region.

In conclusion we reconsidered the work of Bertrand~\textit{et al.}~\cite{b4} 
who treated the schematic Sch\"utte-Da-Providencia model for interacting bosons 
and fermions within the SCRPA scheme. In~\cite{b4} the RPA operator consisted only 
out of one boson and fermions pairs. Here we extended this configuration space and 
included in addition bosons pairs. One of the motivations to do this was to see 
whether the Goldstone theorem which was quite strongly violated in~\cite{b4} is improved. 
It was found that the low-lying mode in the "deformed" zone, \textit{i.e.} symmetry 
broken region, indeed is lowered by $\sim 30\%$ when boson pair terms are added to 
the RPA operator. However, with respect to the first physical state, the position 
of the spurious mode is still too high and one therefore can not say that it decouples 
to a good approximation from the physical spectrum. However, in spite of this somewhat 
disappointing result, the introduction of the extra terms allowed to reproduce very well 
a further excited state of the spectrum in the symmetry conserved phase and additionally the 
quantities which had already been calculated in~\cite{b4} without the boson pair 
operators are still improved. We also calculated further quantities 
as the number of bosons in the ground-state and the fluctuation of the symmetry 
operator. For instance the latter becomes strongly reduced with respect to its mean 
field value. However, also for this quantity a substantial non vanishing value remains. 
All in all we can say that the inclusion of the extra two boson piece to the RPA operator 
allowed to calculate one more state in the spherical region, improve existing results from 
calculations without these terms, and lower the spurious state. However, with respect 
to the latter feature no real breakthrough could be observed and further ideas are needed 
to substantially improve the situation in the symmetry broken region whereas the "spherical"
region seems to be well under control. In this respect this may be similar with other 
approaches treating correlations beyond mean field like coupled cluster theory, 
Jastrow, correlated basis function etc.   

\begin{acknowledgments}
We are specially grateful to Guy Chanfray and Peter Schuck for many interesting 
discussions and stimulating comments. 
We thank Raouf Bennaceur and Jorge Dukelsky for criticism and helpful comments. 
We gratefully acknowledge stimulating discussions with M. Oertel, D. Davesne and H. Hansen.
\end{acknowledgments}

\end{document}